\documentclass[pre, aps, superscriptaddress, showpacs, floatfix, twocolumn]{revtex4-2}
\usepackage[utf8]{inputenc}
\usepackage[T1]{fontenc}
\usepackage[english]{babel}
\usepackage{graphicx}
\usepackage{enumitem}
\usepackage{hyperref}
\usepackage{amsthm}
\usepackage{amssymb}
\usepackage{appendix}
\usepackage{tensor}
\usepackage{physics}
\usepackage{bbold}
\usepackage{amsfonts}
\usepackage{placeins}
\usepackage{xcolor}
\usepackage{mathdots}
\usepackage{numprint}
\usepackage{microtype}
\usepackage{mathtools}
\usepackage{tikz}

\usepackage[normalem]{ulem}

\definecolor{myred}{HTML}{D62728}

\definecolor{mygray1}{HTML}{DDDDDD}
\definecolor{mygray2}{HTML}{CCCCCC}
\definecolor{mygray3}{HTML}{AAAAAA}
\definecolor{mygray4}{HTML}{888888}
\definecolor{mygray5}{HTML}{666666}
\definecolor{mygray6}{HTML}{444444}
\definecolor{mygray7}{HTML}{222222}
\definecolor{mygray8}{HTML}{000000}

\newcommand{\myshadebottom}[4]{
    \begin{scope}
        \shade[left color=#1, right color=#2] [even odd rule] (#3,0) arc (180:360:#4) (#3+0.02,0) arc (180:360:#4-0.02) ;
    \end{scope}
}
\newcommand{\myshadetop}[4]{%
    \begin{scope}
        \shade[left color=#1, right color=#2] [even odd rule] (#3,0) arc (180:0:#4) (#3+0.02,0) arc (180:0:#4-0.02) ;
    \end{scope}
}

\hypersetup{
    colorlinks,
    citecolor=blue,
    filecolor=black,
    linkcolor=blue,
    urlcolor=black
}

\newcommand{\e}[1]{\mathrm{e}^{#1}}

\newcommand{\braketmatrix}[3]{\left \langle #1 \middle| #2 \middle| #3 \right \rangle}
\newcommand{\inlinebraketmatrix}[3]{\langle #1 \vert #2 \vert #3\rangle}

\usepackage[normalem]{ulem}
\usepackage{cancel}

\numberwithin{equation}{section}

\begin{document}

\title{Coherent backscattering and coherent forward scattering effects in variations of the random quantum kicked rotor}

\author{Hugo Thomas}
\affiliation{MajuLab, CNRS-UCA-SU-NUS-NTU International Joint Research Unit, Singapore}
\affiliation{Centre for Quantum Technologies, National University of Singapore, Singapore}
\affiliation{D\'epartement de Physique, \'Ecole Normale Sup\'erieure, Universit\'e PSL, Paris, France}

\author{Julien H\'ebraud}
\affiliation{Laboratoire de Physique Th\'eorique, Universit\'e de Toulouse, CNRS, UPS, France}

\author{Bertrand Georgeot}
\affiliation{Laboratoire de Physique Th\'eorique, Universit\'e de Toulouse, CNRS, UPS, France}

\author{Gabriel Lemari\'e}
\affiliation{Laboratoire de Physique Th\'eorique, Universit\'e de Toulouse, CNRS, UPS, France}
\affiliation{MajuLab, CNRS-UCA-SU-NUS-NTU International Joint Research Unit, Singapore}
\affiliation{Centre for Quantum Technologies, National University of Singapore, Singapore}

\author{Christian Miniatura}
\affiliation{Universit\'e  C\^ote  d'Azur,   CNRS,   INPHYNI,   Nice,   France}
\affiliation{Centre for Quantum Technologies, National University of Singapore, Singapore}

\author{Olivier Giraud}
\affiliation{MajuLab, CNRS-UCA-SU-NUS-NTU International Joint Research Unit, Singapore}
\affiliation{Centre for Quantum Technologies, National University of Singapore, Singapore}
\affiliation{Universit\'e Paris-Saclay, CNRS, LPTMS, 91405 Orsay, France}

\date{March 4, 2025}

\begin{abstract}
We investigate coherent multiple scattering effects in the random quantum kicked rotor model. By changing the starting time of the Floquet period, two new classes of models can be introduced that exhibit similar interference structures. For one of the two classes, these structures appear on top of a non-trivial background, which we describe in detail. Its origin is identified and an approximate analytical formula is proposed. 
The asymptotic contrast profile, as well as the height of the coherent backscattering and the coherent forward scattering peaks, are also analyzed.
Our findings are relevant to provide an interpretation of cold atom experiments aiming at observing such interference effects with matter waves.
\end{abstract}

\maketitle

\setcounter{tocdepth}{1}
\makeatletter
\def\l@subsubsection#1#2{}
\makeatother

\section{Introduction}
\label{sec:introduction}

Coherent backscattering (CBS) is arguably the emblematic signature interference effect associated with coherent linear wave propagation in disordered media. As it is an ubiquitous effect, it has been observed using very different classical waves and very different disordered media in the weak (semi-classical) localization regime: optical, acoustic and seismic~\cite{albada_observation_1985, wolf_weak_1985, bayer_weak_1993, wiersma_1995, labeyrie_coherent_1999, larose_weak_2004, cobus_2017}. As a matter of fact, CBS can be demonstrated with quite simple optical set-ups \cite{corey_1995}. Its physical origin can be traced back to disorder-immune constructive two-wave interference terms involving scattered waves contra-propagating along the same scattering paths~\cite{akkermans_coherent_1986}.

CBS has also been observed with spinless atomic matter waves \cite{jendrzejewski_cbs_2012, labeyrie_cbs_2012}. 
In this setting, atoms are prepared and launched 
with an initial momentum $p_0$ within a spatially disordered potential, usually produced with a speckle light field. Because successive scattering events randomize wave propagation, the disorder-averaged momentum distribution at large enough times is expected to become isotropic and to fully spread over the available energy shells. This momentum distribution is called the (incoherent) diffusive background. For time-reversal invariant systems however, a CBS interference peak appears atop this diffusive background at momentum $-p_0$ after a time of the order of the scattering mean free time \cite{cherroret_cbs_2012}.

In the weak localization regime, wave transport is only weakly affected by interference. Anderson understood that disorder could in fact impact wave transport even more dramatically~\cite{anderson_absence_1958} with profound consequences, as decribed by the scaling theory of localization~\cite{gang4_1979}. In this strong localization regime,  destructive quantum interferences simply bring diffusion to a complete stop and the medium behaves like an insulator. Since then, Anderson localization has been observed and studied in several systems~\cite{billy_2008, roati_2008, hu_2008, lahini_2008, jendr_2012, segev_2013, barbosa_2024} but remains elusive for light in 3D~\cite{skipetrov_2016}. 

Recently, in the context of matter waves, it has been discovered that Anderson localization triggers the emergence of an interference peak twinning the CBS peak in the long time limit \cite{karpiuk_coherent_2012, lee_dynamics_2014, ghosh_2D_2014}. The appearance of this novel interference peak, dubbed the coherent forward scattering (CFS) effect, is a marker of localization, even embodying profound features of the 3D Anderson transitions like critical exponents and multifractal dimensions~\cite{ghosh_cfs_2017, ehsan_2024} \footnote{Note that the CBS peak can also reveal the Anderson transition \cite{ghosh_cbs_2015}}. It appears at a time of the order of the localization time (Heisenberg time associated to a localization volume) $\tau_H$ usually larger than the timescale associated to the emergence of the CBS peak.

\begin{figure*}[!t]
    \centering
    \includegraphics[width=.49\linewidth]{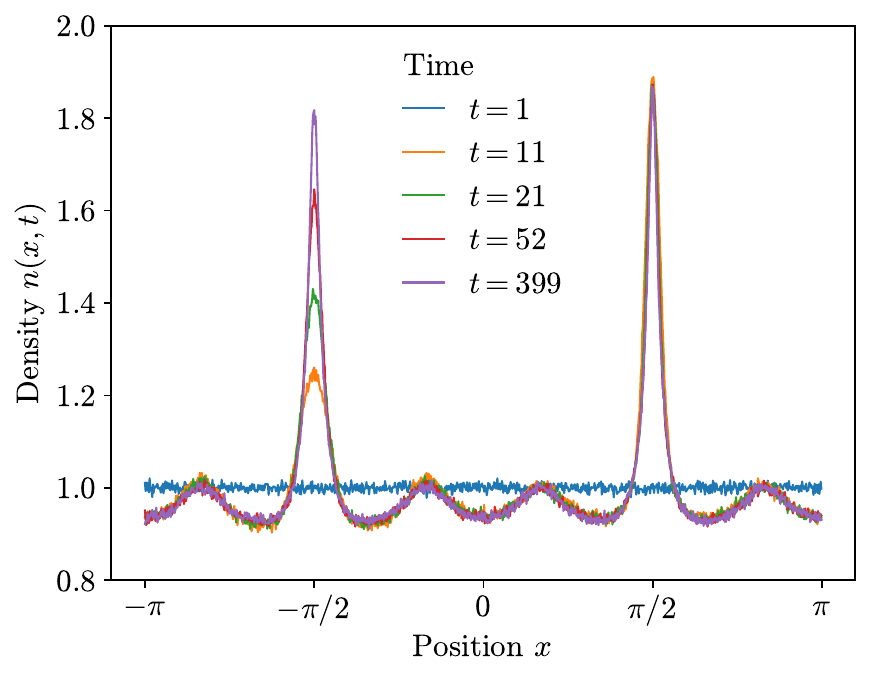}
    \includegraphics[width=.49\linewidth]{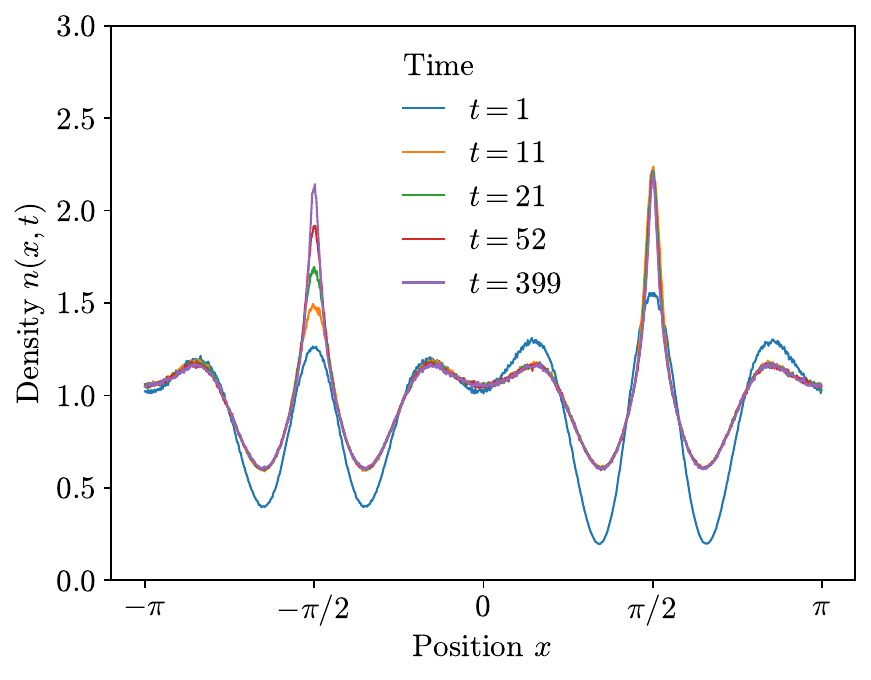}
    \caption{\textbf{RQKR Dynamics} - Stroboscopic Floquet evolution of the spatial probability density $n(x,t)$ at different times obtained under $U_0 = U_xU_p$ (left panel) and $U_{1/2}=\sqrt{U_p}U_x\sqrt{U_p}$ (right panel) for an initial condition peaked at $x_0 = -\pi/2$. The parameters are $N = 1024$, $K=20$, $\hbar_e = 2.89$ and the random (uncorrelated) phases $\alpha_p$ are drawn uniformly over $[ 0, 2 \pi [$. The average is done over $\numprint{20000}$ phase disorder configurations. The evolution under the operator $V_{1/2}$ is not shown here as it gives the same spatial distribution as $U_0$.}
    \label{fig:qkr_evolution}
\end{figure*}

Noticeably, similar phenomena can be observed and studied using deterministic systems subject to time-periodic driving. In this context, disorder is produced dynamically, and the time evolution is described stroboscopically by iterating a unitary quantum (Floquet) map at discrete time steps. 
As the quantum map incorporates the effect of dynamical disorder over one time step, such systems provide simple yet powerful models to investigate properties of dynamically-disordered quantum systems at large times. This is the reason why they have been extensively studied in the context of quantum chaos~\cite{haake_2010}. A paradigmatic Floquet model is the quantum kicked rotor (QKR). Here, the system is periodically kicked with a position-dependent strength and evolves freely between two consecutive kicks~\cite{santhanam_quantum_2022}. This system exhibits dynamical localization, a phenomenon akin to Anderson localization in 1D~\cite{fishman_chaos_1982, grempel_1984}. Dynamical localization was observed with cold atoms~\cite{moore_1994, moore1995atom, raizen_experimental_1995}, and, using incommensurate driving frequencies~\cite{casati_1989}, used to tackle experimentally the 3D Anderson localization transition, its critical state~\cite{chabe_2008, lemarie_2010, akridas_2019} and the so-called quantum boomerang effect~\cite{prat_2019, tessieri_2021, sajjad_2022, janarek_2022}.

Such Floquet systems turn out to be convenient experimental platforms to observe and study the CBS and CFS interference phenomena too. However, since disorder is imprinted in momentum space by the (quasi-)random free evolution phases, the CBS and CFS peaks emerge now in the disorder-averaged spatial distribution of the system~\cite{lemarie_coherent_2017}. An advantage of the QKR model is that it can be adapted easily to explore different symmetry classes (orthogonal, unitary and symplectic) and regimes (localized, critical and delocalized)~\cite{scharf_1989, hainaut_sym_2018, martinez_coherent_2021, martinez_coherent_2023}.

At the heart of the present paper is the following observation. In periodically kicked systems, the stroboscopic approach describes the state of the system at each period of time by repeated action of the corresponding Floquet operator $U$ on an initial state. One would intuitively expect that the universal signatures of localization discussed above would be insensitive to the initial time of the stroboscopic observations: whether we start the dynamic evolution by a kick or by a free evolution should lead to similar results. Quite unexpectedly, this is not the case. We illustrate this in Fig.~\ref{fig:qkr_evolution}, which displays the disorder-averaged spatial probability density $n(x, x_0, t) = \overline{|\psi(x,t)|^2} = \overline{|\bra{x}U^t\ket{x_0}|^2}$ obtained at several discrete times $t$ when the system is prepared initially in some position state $\ket{x_0}$. The CBS and CFS peak structures appear at positions $-x_0$ and $x_0$, respectively. While CBS has fully developed after a few kicks, the CFS takes a longer time to fully emerge.
The left panel of Fig.~\ref{fig:qkr_evolution} shows the results obtained with the usual QKR Floquet protocol, where during each period of time the system evolves freely, and is subjected to a kick at the end of the period. On the other hand, the right panel shows the results obtained when the system first evolves freely over half a period of time, then is kicked, and finally completes another free evolution over half a period of time. As one can see, the resulting spatial profiles of the density are totally different, with high spatial oscillations in the second case, and seemingly different heights and contrasts of the CBS and CFS peaks at $-x_0$ and $x_0$.

The goal of the present paper is to analyze and clarify the origin of these different behaviors. One may think that these additional oscillations are due to interference effects beyond CBS and CFS. On the contrary, we show that the oscillations that manifest themselves for different choices of the starting time of the Floquet sequence are due to an oscillatory non-interferential background; after subtracting this contribution, we get spatial peak structures which closely match one another. In order to experimentally observe interferometric signatures such as CBS or CFS  it is necessary to know in detail the background from which they emerge. Our work is thus important in order to characterize such signatures in cold atom experimental realizations of the QKR. 

After defining our models in Section~\ref{sec:models}, we calculate in Section~\ref{sec:coherent_scattering} the density profile of an iterated quantum state at large time and find analytic formulas for the background. A discussion is given in Section~\ref{sec:conclusion}.

\section{Models}
\label{sec:models}

In this Section, we introduce the random quantum kicked rotor (RQKR) model and two of its variants. For each of them, we analyse in detail the behavior of a wavefunction in position space under Floquet iteration.

\subsection{Definition of the RQKR models}
\label{subsec:mdl_presentation}

\subsubsection{Hamiltonian and Floquet evolution operators}
\label{subsubsec:mdl_hamiltonian}

In dimensionless units, the QKR Hamiltonian takes the form
\begin{equation}
    H(t) = \dfrac{\hat{p}^2}{2} - K \cos \hat{x} \sum_{n=-\infty}^\infty \delta (t-n),
\label{eq:qkr_hamiltonian}
\end{equation}
where $[\hat{x},\hat{p}]=i\hbar_e$. Here, $K$ and $\hbar_e$ are the dimensionless kick strength and effective Planck's constant, respectively. The evolution of a system subjected to this Hamiltonian consists of a free evolution interrupted at regular intervals $t_n=nT$ ($n\in \mathbb{Z}$ and $T=1$ here) by position-dependent kicks of amplitude $K$. Following the Floquet approach, the dynamics can be studied stroboscopically at each multiple of the time period $T=1$. This leaves open the choice of the start time $t_0 \in \left[ 0, 1 \right]$ of the periodic sequence used for the stroboscopic description, that is, how to distribute the free evolutions around the kick. This choice leads to a continuous family of unitary Floquet operators parametrized by $t_0$. For a given choice of $t_0$, we have $U(t_0 + n, t_0) = U_{t_0}^n$ where $U_{t_0} = U(t_0+1,t_0)$. Using the time evolution chain rule for Hamiltonian \eqref{eq:qkr_hamiltonian}, we have 
\begin{equation}
    U_{t_0} = U^{t_0}_p \,U_{x}\,U^{1-t_0}_{p} = U^{t_0}_p \,(U_{x}U_p)\,U^{-t_0}_{p}\,,
\label{eq:qkrt0_evolution_operator}
\end{equation}
where $U_p=e^{-i \frac{\hat{p}^2}{2\hbar_e}}$ describes the free evolution over one period and $U_x=e^{i \frac{K}{\hbar_e} \cos \hat{x}}$ describes the instant evolution under the $\delta$-kick.

Trivially, Eq.~\eqref{eq:qkrt0_evolution_operator} shows that all Floquet operators $U_{t_0}$ are unitarily related to $U_xU_p$ and thus have the same Floquet quasi-energy spectrum than $U_xU_p$. 
However, their respective corresponding eigenkets are unitarily related to those of $U_xU_p$ by $U^{t_0}_p$. Since disorder in QKR systems is imprinted through free dynamical phases, the difference in free evolutions, captured by $U^{t_0}_p$, will impact wave function amplitudes and thus, possibly, their statistical properties. As a direct consequence, physical observables computed with them should be affected too, as will be seen below.

\subsubsection{Usual QKR and its RQKR variant }

\begin{figure}
    \centering

        \centering
        \begin{tikzpicture}[scale=1]
        \draw[->, semithick] (-0.5,0) -- (2.7,0) node[below] {\textit{t}};
        \draw[semithick] (0,-0.3) -- (0,0.3);
        \draw[semithick] (2,-0.3) -- (2,0.3);
        \draw[->, semithick] (0,-0.6) -- (2,-0.6) node[midway, below=2pt] {$U_0$};
        \draw[semithick, -] (0,0.5) to[out=80, in=180] (0.95,1.2);
        \draw[semithick, ->] (0.95,1.2) to[out=0, in=100] (1.9,0.5);
        \node at (0.95,1.6) {$U_p$};
        \draw[semithick, myred, ->] plot[domain=-0.97:1] ({2 + 0.1*\x}, {0.5 + 0.7*sqrt(1 - \x*\x)});
        \node[myred] at (2,1.6) {$U_x$};
        \node at (-0.6, 1.6) {(a)};
    \end{tikzpicture}
    \hspace{0.3cm}
    \begin{tikzpicture}[scale=1]
        \draw[->, semithick] (-0.1,0) -- (3.4,0) node[below] {\textit{t}};
        \draw[semithick] (0,-0.3) -- (0,0.3);
        \draw[semithick] (2,-0.3) -- (2,0.3);
        \draw[->, semithick] (1,-0.6) -- (3,-0.6) node[midway, below=2pt] {$U_{1/2}$};
        \draw[semithick, ->, samples=100] plot[domain=-1:1] ({1.45 + 0.45*\x}, {0.5 + 0.7*sqrt(1 - \x*\x)});
        \node at (1.35,1.6) {$\sqrt{U_p}$};
        \draw[semithick, ->, samples=100] plot[domain=-0.93:1] ({2.55 + 0.45*\x}, {0.5 + 0.7*sqrt(1 - \x*\x)});
        \node at (2.55,1.6) {$\sqrt{U_p}$};
        \draw[semithick, myred, ->, samples=100] plot[domain=-0.93:1] ({2 + 0.1*\x}, {0.5 + 0.7*sqrt(1 - \x*\x)});
        \node[myred] at (2,1.6) {$U_x$};
        \node at (-0.2, 1.6) {(b)};
    \end{tikzpicture}
    
    \begin{tikzpicture}[scale=1]
        \draw[->, semithick] (-0.7,0) -- (2.7,0) node[below] {\textit{t}};
        \draw[semithick] (0,-0.3) -- (0,0.3);
        \draw[semithick] (2,-0.3) -- (2,0.3);
        \draw[->, semithick] (0,-0.6) -- (2,-0.6) node[midway, below=2pt] {$V_{1/2}$};
        \draw[semithick, -] (0.12,0.65) to[out=65, in=180] (1,1.2);
        \draw[semithick, ->] (1,1.2) to[out=0, in=100] (1.9,0.5);
        \node at (1,1.6) {$U_p$};
        \draw[semithick, myred, ->] plot[domain=-1:1] ({0.07*\x}, {0.5 + 0.7*sqrt(1 - \x*\x)});
        \node[myred] at (0,1.6) {$\sqrt{U_x}$};
        \draw[semithick, myred, ->] plot[domain=-0.98:1] ({2 + 0.07*\x}, {0.5 + 0.7*sqrt(1 - \x*\x)});
        \node[myred] at (2,1.6) {$\sqrt{U_x}$};
        \node at (-0.8, 1.6) {(c)};
    \end{tikzpicture}
    
    \caption{\textbf{Evolution operators} - Three different Floquet evolution operators are considered. Their schematic representation is given above on a timeline where the ticks represent the kicks that occur at integer times. (a) The evolution operator $U_0=U_xU_p$ corresponds to a free evolution $U_p$ during one period of time $T=1$, followed by a full kick $U_x$ of strength $K$. (b) The evolution operator $U_{1/2}=\sqrt{U_p} \, U_x \,\sqrt{U_p}$ corresponds to two free evolution $\sqrt{U_p}$ lasting half a period of time and separated by a full kick $U_x$. (c) The evolution operator $V_{1/2}$ corresponds to a free evolution $U_p$ during one period of time boxed by two kicks $\sqrt{U_x}$ of half strength $K/2$. 
    }
    \label{fig:floquet_diagram}
\end{figure}
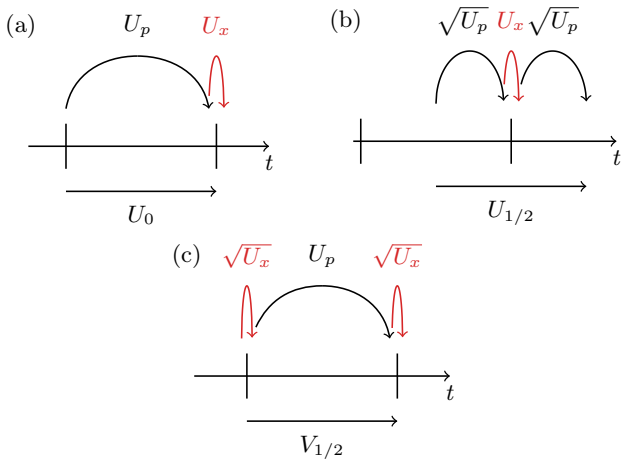

The QKR model is obtained for $t_0=0$, that is, by  
having the full free evolution during a time $T$ before a kick (see Fig.~\ref{fig:floquet_diagram}a). The QKR Floquet evolution operator over one period is thus
\begin{equation}
    U_0 = U_x \, U_p = e^{i \frac{K}{\hbar_e} \cos \hat{x}} \, e^{-i \frac{\hat{p}^2}{2 \hbar_e}}\,.
\label{eq:qkr0_evolution_operator}
\end{equation}
Since the potential is periodic in space, Bloch theorem implies that eigenfunctions satisfy $\phi_n(x+2\pi) = \exp (i\beta x) \, \phi_n(x)$ where the quasi-momentum $\beta \in [0, 1]$. This also means that the convenient representation of momenta states is $\ket{p} = \ket{(l+\beta)\hbar_e} \equiv \ket{l, \beta}$, with $l \in \mathbb{Z}$. The case $\beta =0$ describes a kicked pendulum for which the space variable identifies with an angle. The matrix elements of $U_0$ in momentum space are given by
\begin{equation}
    \bra{l', \beta'} U_0 \ket{l, \beta} = e^{-i \varepsilon_p} \, i^{l'-l} J_{l'-l}(K/ \hbar_e) \delta(\beta' - \beta),
\label{eq:qkr0_evolution_operator_momentum}
\end{equation}
where $J_n(x)$ is the $n$th Bessel function and $\varepsilon_p = (l+\beta)^2 \hbar_e /2$ is the dynamical phase induced by the free evolution operator over one period, $U_p$, in momentum space. A consequence of Eq.~\eqref{eq:qkr0_evolution_operator_momentum} is that the evolution operator $U_0$ conserves the pseudo-momentum $\beta$, which is thus a constant of motion. In numerical computations, one generally truncates the possible $l$ values to $|l| \leq N$, where $N$ plays the role of the system size in momentum space and is chosen large enough to capture the relevant physical behavior being targeted. Equation~\eqref{eq:qkr0_evolution_operator_momentum} also shows that the kick operator $U_x$ induces a hopping in the $l$-lattice which is restricted to a range of order $K/\hbar_e$ by the exponential decrease of the Bessel function~\cite{abramowitz1968handbook}. As a consequence, $\ell_s = K/\hbar_e$ plays the role of an effective scattering mean free path in momentum space.

The dynamical phases $\varepsilon_p$ in \eqref{eq:qkr0_evolution_operator_momentum}, though completely deterministic, turn out to be pseudo-random numbers uniformly distributed over $\left[ 0, 2 \pi \right]$ when $p$ varies, as soon as $\hbar_e$ is incommensurate with $2 \pi$. A practical drawback of the QKR is that disorder averaging is however not available, as the model is completely deterministic. To unveil its quasi-random nature, one would have to resort to averaging physical observables over well-chosen initial states. This also comes with some drawback when discussing symmetries, in particular time reversal invariance (see Section~\ref{subsec:symmetries}), as it flips the sign of momentum and thus the sign of $\beta$. Since the QKR evolution preserves $\beta$, one would mix evolutions with different symmetry properties.
To circumvent these practical drawback, following \cite{fishman_chaos_1982, griniasty1988localization, brenner1992pseudo}, one defines the random quantum kicked rotor (RQKR) model by replacing the deterministic kinetic operator $\hat{p}^2/(2\hbar_e)$ by a random one $\alpha(\hat{p}) = \sum_p \,\alpha_p \, \ket{p}\!\bra{p}$, diagonal in momentum state. Equivalently, this amounts to replacing the dynamical phases $\varepsilon_p$ in Eq.~\eqref{eq:qkr0_evolution_operator_momentum} by truly random phases $\alpha_p$. Disorder average is then an average over these random phases $\alpha_p$. Another advantage of this approach is that now different types of phase disorder can be considered, like an Anderson-like partial phase disorder for example~\cite{haldar_2023}. For the rest of this paper, we consider the RQKR and $U_p = \exp[-i\alpha(\hat{p})]$. Since $U_x$ still conserves $\beta$, and without any real loss of generality, we stick here to the case $\beta=0$. From now on, $p$ takes on integer values. Note that disorder is introduced in momentum space, and therefore the physical behaviors in position and momentum space are interchanged compared to the standard Anderson picture described in the introduction. This means in particular that the CBS and CFS peaks will emerge in position space.

The model $U_0$ corresponds to a stroboscopic evolution starting at $t_0=0$. We now consider two different choices for the starting point of the evolution, which will yield two different evolution operators.


\subsubsection{Time-symmetrized QKR and RQKR}
\label{subsubsec:mdl_symmetrization}
For this model, we equally distribute the free evolutions around the kick (be it for the QKR or the RQKR). This is achieved by choosing $t_0=1/2$ (see Fig.~\ref{fig:floquet_diagram}b) and we have
\begin{equation}
    U_{1/2} = 
    \sqrt{U_p}\,U_x\,\sqrt{U_p}\,.
\label{eq:qkr_tau_evolution_operator}
\end{equation}
The unitary equivalence between $U_{1/2}$ and $U_0$ is given by the random operator $\sqrt{U_p} \equiv e^{-i\alpha(\hat{p})/2}$.

\subsubsection{Kick-symmetrized QKR and RQKR}
For this model, we introduce a new evolution scenario where two $\delta$-kicks of amplitude $Kr$ and $K(1-r)$ are bundled together and we look at the wavefunction "in between them". The corresponding evolution operator reads 
\begin{equation}
    V_{r} = U^{r}_x \, U_p \, U^{1-r}_x = U^{r-1}_x \, (U_xU_p) \, U^{1-r}_x\,.
\label{eq:qkr_half_kick_evolution_operator}
\end{equation}
In analogy with Eq.~\eqref{eq:qkrt0_evolution_operator}, it is as if the kick had been distributed around the full free evolution. This prescription introduces a continuous family of evolution operators parametrized by $r$. Being unitarily equivalent to $U_0$, all $V_r$ and $U_{t_0}$ have the same Floquet quasi-energy spectrum irrespective of the actual value of $r$ and $t_0$. Furthermore, the eigenkets of $V_r$ are unitarily related to those of $U_0$ by $U_x^{r-1}$. There is however a huge difference here as $U_x$ does not contribute any randomness to the evolution. This has dramatic consequences when comparing the physical properties of the system evolving under $U_{t_0}$ or under $V_r$.

In full analogy with Eq.~\eqref{eq:qkr_tau_evolution_operator}, the kick-symmetrized case is obtained for $r=1/2$ (see Fig.~\ref{fig:floquet_diagram}c) and reads
\begin{equation}
    V_{1/2} = 
    \sqrt{U_x} \, U_p \, \sqrt{U_x}
\label{eq:qkr1/2_half_kick_evolution_operator}
\end{equation}
where $\sqrt{U_x} \equiv e^{i \frac{K}{2\hbar_e}\cos \hat{x}}$. The unitary equivalence between $V_{1/2}$ and $U_0$ is given by the random operator $\sqrt{U^{\dag}_x} \equiv e^{-i \frac{K}{2\hbar_e}\cos \hat{x}}$.

In the rest of the Paper, we mainly focus on the dynamical properties of the system evolving under either the time-symmetric  $U_{1/2}$ or the kick-symmetric $V_{1/2}$ evolution operators.

\subsection{Discrete Symmetries}
\label{subsec:symmetries}

The discrete symmetries relevant to the RQKR and its variants are parity $\mathsf{P}$, time reversal $\mathsf{T}$, and their combination $\mathsf{PT}$. Importantly, $\mathsf{T}$ is anti-unitary~\cite{sigwarth_2022} and, as a consequence, $\mathsf{PT}$ too. Both realize time reversal operators in appropriate bases of the Hilbert space, see Appendix~\ref{appconsequences}. For spinless systems like here, $\mathsf{T}$ is simply the complex conjugation operator $\hat{K}_x$ defined in position space, while $\mathsf{PT}$ is the complex conjugation operator $\hat{K}_p$ defined in momentum space, see Appendix~\ref{appconsequences}.

It turns out that both models $U_{1/2}$ and $V_{1/2}$ are $\mathsf{PT}$-symmetric and are thus time reversal invariant in the momentum basis. 
Indeed, it is easy to check that

\begin{equation}
\begin{aligned}
\hat{K}_p U_{1/2} \hat{K}_p & = \hat{K}_p e^{-i \alpha(\hat{p})/2} \, e^{i \frac{K}{\hbar_e} \cos\hat{x}} \, e^{-i \alpha(\hat{p})/2} \hat{K}_p \\
& = e^{i\alpha(\hat{p})/2} \, e^{-i \frac{K}{\hbar_e} \cos\hat{x}} \, e^{i \alpha(\hat{p})/2} \\
& = U_{1/2}^\dagger\,.
\end{aligned}
\label{eq:qkr_time_reversal_Utau}
\end{equation}
Similarly, we have $\hat{K}_p V_{1/2} \hat{K}_p= V_{1/2}^\dagger$. As discussed in Appendix~\ref{appconsequences}, this $\mathsf{PT}$-symmetry entails that eigenstates $\ket{\varphi_a}$ of $U_{1/2}$ can be chosen such that $\varphi_a(p)$ is real and $\varphi_a^*(-x)=\varphi_a(x)$. The same type of conclusion holds for eigenstates of $V_{1/2}$.

One can easily check that the initial RQKR model $U_0$ is not $\mathsf{PT}$-symmetric since $\hat{K}_p \, U_0 \, \hat{K}_p = U_x^\dag U_p^\dag \neq U_0^\dag$. However, it is important to note that $U_0$ still retains time reversal invariance. Indeed, introducing the anti-linear operator $\Theta = U_x^\dag \hat{K}_p$, we see that $\Theta \, U_0 \, \Theta = U_0^\dag$. Thus, it is $\Theta$, rather than $\hat{K}_p$ alone, which plays the role of a time reversal operator for $U_0$. In this case, one can show that $\varphi^*_a(-x) = e^{iK\cos x /\hbar_e} \varphi_a(x)$.

In view of the above, it can appear more natural, when investigating the role of symmetries in the quantum kicked rotor, to consider $U_{1/2}$ and $V_{1/2}$ rather than $U_0$ for which time reversal acts through a less trivial operator. In the next Section, we explore the consequences of such a choice.

\section{Coherent scattering effects}
\label{sec:coherent_scattering}

We now compare the different stroboscopic evolutions obtained when $U_0$, $U_{1/2}$ or $V_{1/2}$ repeatedly act on a given initial state $\ket{\psi_0}$. For the QKR model, Floquet eigenstates of $U_0$ are exponentially localized in momentum space  with a localization length $\xi$ scaling like~\cite{lemarie_coherent_2017}
\begin{equation}
    \xi \sim \frac{K^2}{4 \hbar_e^2}\,.
\label{eq:qkr_localization_length}
\end{equation}
When $\xi$ is of the order of the system size $N$ (that is, when the kick amplitude $K$ is large enough), the eigenstates extend almost over the entire momentum space and the system retains metallic features. On the other hand, when $\xi\ll N$ (that is, when $K$ is small enough), the eigenstates are well localized and the system is truly insulating. It is this latter regime that we consider in the following.

\subsection{Spatial probability density}

As localization takes place in momentum space, the CBS and CFS interference peaks appear in position space. We thus compute the spatial probability density associated with a state initially localized in position space, $\ket{\psi_0} = \ket{x_0}$ with $x_0 \in \left[- \pi, \pi \right]$.

After $t \in \mathbb{N}$ kicks, the state is $\ket{\psi (t)} = U^t \ket{x_0}$, where $U$ is the Floquet operator under consideration. We want to describe the behavior of the disorder-averaged probability density $n(x, x_0, t) = \overline{|\psi(x,t)|^2}$, where $\psi(x, t) = \braket{x}{\psi(t)}$ are the components of $\ket{\psi (t)}$ in position basis. The density is normalized over the unit cell $x \in \left[- \pi, \pi \right]$ by
\begin{equation}
    \dfrac{1}{N} \sum_x n(x,x_0,t) = 1 \underset{N\rightarrow \infty}{\longrightarrow} \int_{-\pi}^\pi \dfrac{\text{d}x}{2 \pi} \; n(x, x_0, t) = 1
\label{eq:qkr_density_normalization}
\end{equation}
(see Appendix~\ref{sec:appendix:normalization} for the normalization conditions and limiting behavior $N\to\infty$). We have 
\begin{equation}
    \bra{x}V_{1/2}^t\ket{x_0} =  e^{i \frac{K}{2 \hbar_e}(\cos x_0 - \cos x )} 
    \bra{x}U_0^t\ket{x_0}\,,
\label{eq:equiv_0_pos_nondiag}
\end{equation}
and thus the matrix elements of $V^t_{1/2}$ and $U^t_0$ in position space are equal up to a space-dependent phase factor. Therefore, the  probability densities at time $t$, $n(x, x_0, t) = \overline{|\inlinebraketmatrix{x}{U^t}{x_0}|^2}$, are exactly the same for the two models. 

In stark contrast, we have $\bra{x}U_{1/2}^t\ket{x_0} =  
    \bra{x}\sqrt{U_p}\,U_0^t\,\sqrt{U_p}^\dag\ket{x_0}$. Since $U_p$ is not diagonal in position space, the matrix elements $\bra{x}U_{1/2}^t\ket{x_0}$ and $\bra{x}U_0^t\ket{x_0}$ are no longer simply proportional. As a consequence, the spatial probabilities computed for $U_{1/2}$ and for $U_0$ are no longer the same and display striking differences, as shown in Fig.~\ref{fig:qkr_evolution}. Because the dynamics is time-reversal invariant (see Section~\ref{subsec:symmetries}), a fully-contrasted CBS peak appears at position $x_0$ for both models, right after the first kick for $U_0$ and after a few kicks for $U_{1/2}$. For both models, a CFS peak gradually appears at position $x_0$ and fully develops over identical and much longer time scales than CBS. Noticeably, both models present oscillations in the steady state profile reached at large time. However, the standard model $U_0$ has a uniform profile at $t=1$ while the first iteration of  $U_{1/2}$ reveals important oscillations in space. Additionally, these large oscillations survive at later times. Our aim is now to explain these features.

\subsection{Modal decomposition of the spatial density}
\label{subsec:bgnd_formulas}

To understand the differences observed in Fig.~\ref{fig:qkr_evolution} between the two disorder-averaged spatial densities, let us decompose each of them over the Floquet eigenbasis $\left\{\ket{\varphi_n} \right\}_n$ and eigenphases $\left\{\omega_n \right\}_n$ associated with their respective unitary Floquet operator $U=U_0$ or $U=U_{1/2}$. Using $U\ket{\varphi_n}=e^{-i\omega_n}\ket{\varphi_n}$, we find
\begin{equation}
    n(x, x_0, t) = \overline{\sum_{n, m} e^{-i (\omega_n -\omega_m) t} \varphi_n(x) \varphi_m^*(x) \varphi_m(x_0) \varphi_n^*(x_0)}\,.
\label{eq:probability_density}
\end{equation}

The density $n(x, x_0, t)$ can be further decomposed as
\begin{equation}
\label{decompn}
 n(x, x_0,t)  = n_B(x, x_0,t)+n_I(x, x_0,t)  \,,
\end{equation}
where $n_B(x, x_0,t)$ denotes the background and $n_I(x, x_0, t)$ contains all interference contributions.
In the long-time limit the diagonal approximation ($n=m$) can be performed in Eq.~\eqref{eq:probability_density} and the density converges to a steady-state spatial distribution $n_S(x, x_0) = \lim_{t\rightarrow \infty} n(x, x_0, t)$ that reads
\begin{align}
    \label{eq:qkr_steady_state1}
    n_S(x, x_0) & =  \overline{\sum_n |\varphi_n(x)|^2 |\varphi_n(x_0)|^2} \\
    \label{eq:qkr_steady_state2}
    & = N\int_{-\pi}^\pi \text{d}\omega \, \overline{\dfrac{A(x, \omega) A(x_0, \omega)}{\nu(\omega)}}\,,
\end{align}
where the density of states $\nu(\omega)$ and the spectral function $A(x,\omega)$ are defined by
\begin{align}
 \label{eq:defnu}
 \nu (\omega) & = \dfrac{1}{N} \overline{\sum_n \delta (\omega - \omega_n)}\,, \\
   A(x, \omega) & = \dfrac{1}{N} \sum_n \delta (\omega - \omega_n) |\varphi_n(x)|^2\,.
\label{eq:qkr_spectral_function}
\end{align}
At this point, we remind the reader that $U_0$ and $U_{1/2}$ have the same quasi-energy spectrum but different eigenfunction amplitudes in position space. This is because their eigenkets are unitarily related by $\sqrt{U_p}$, which is not diagonal in position space. As a consequence the density of states $\nu(\omega)$ is the same for both $U_0$ and $U_{1/2}$ models while their respective spectral functions in position space do differ, entailing different steady-state spatial distributions. Note, however, that their spectral functions in momentum space would be exactly the same as $\sqrt{U_p}$ is diagonal in $p$-space, leading to identical steady-state momentum distributions. Finally, we remark that $n_S(x,x_0)$ is even in $x$ and $x_0$. This is because $U_0$ is $\Theta$-invariant and $U_{1/2}$ $\mathsf{PT}$-invariant, from which we can easily show, for both models, that $|\varphi_n(-x)|^2 = |\varphi_n(x)|^2$.

As shown by the normalization conditions  \eqref{eq:qkr_normalization_A_omega}--\eqref{eq:qkr_normalization_A_nu_x}, the disorder-averaged spectral function $\bar{A}(x, \omega)$ can be interpreted as the classical probability distribution for the system to have energy $\omega$ when it is at position $x$. By the same token, $\bar{A}(x, \omega)/(2\pi\nu(\omega))$ is interpreted as the classical probability for the system to be at position $x$ when it has energy $\omega$. 
Using probability chain rules, the classical probability for the system to be at position $x$ when it started at position $x_0$ reads
\begin{equation}
    n_B(x, x_0) = N\int_{-\pi}^\pi \text{d}\omega \, \dfrac{\bar{A}(x, \omega) \bar{A}(x_0, \omega)}{\nu(\omega)}\,,
\label{eq:qkr_background}
\end{equation}
which corresponds to a summation over all (classical) ways of going from $x_0$ to $x$ at different energies. Comparing Eqs. \eqref{eq:qkr_steady_state2} and \eqref{eq:qkr_background}, we see that the classical background corresponds to the contributions to $n_S(x, x_0)$ for which the components of the spectral function corresponding to different positions are decoupled. Note that $n_B(x,x_0)$, like $n_S(x,x_0)$, is even in $x$ and $x_0$.

Importantly, in the $N\to\infty$ limit, both $n_S(x,x_0)$ and $n_B(x,x_0)$ obey Eq.~\eqref{eq:qkr_density_normalization} and are normalized to unity. The steady state $n_S(x, x_0)$ in Eq.~\eqref{eq:qkr_steady_state2} can then be decomposed as
\begin{equation}
\label{decompsteady}
 n_S(x, x_0)  = n_B(x, x_0)+n_I(x, x_0),
\end{equation}
where $n_B(x, x_0)$ defines the interference-free classical background and $n_I(x, x_0)$ encapsulates the interference contributions. The latter is normalized to zero,
\begin{equation}
  \dfrac{1}{N} \sum_x n_I(x, x_0)= 0 \underset{N\rightarrow \infty}{\longrightarrow} \int_{-\pi}^\pi \dfrac{\text{d}x}{2 \pi} \; n_I(x, x_0) = 0\,,
\label{eq:qkr_interf_density_normalization}
\end{equation}
and is even in $x$ and $x_0$.

The disorder-averaged spectral function $\bar{A}(x, \omega)$ can be calculated as follows.
Using the Floquet eigenbasis of $U$, we have 
\begin{align}
 \inlinebraketmatrix{x}{\overline{U^t}}{x} & = \overline{\sum_n e^{-i \omega_n t} |\varphi_n(x)|^2} \\
& = N\int_{-\pi}^{\pi} \text{d}\omega \, e^{-i \omega t} \bar{A}(x, \omega).
\label{eq:qkr_ballistic_decay}
\end{align}
By inverse Fourier transform, we find
\begin{equation}
\bar{A}(x, \omega) = \frac{1}{2\pi N} \sum_{t=-\infty}^{\infty} e^{i \omega t} \inlinebraketmatrix{x}{\overline{U^t}}{x}\,.
\label{eq:qkr_spectral_function_from_ballistic}
\end{equation}
Note that $\inlinebraketmatrix{x}{\overline{U^{-t}}}{x} = \inlinebraketmatrix{x}{\overline{U^t}}{x}^*$, where the star denotes complex conjugation. This implies that $\bar{A}(x, \omega)$ is indeed real as it should. The background is then numerically obtained via Eq.~\eqref{eq:qkr_background}, as we show below.

\subsection{Background for $U_0$ and $V_{1/2}$}
\label{subsec:bgnd_standard}

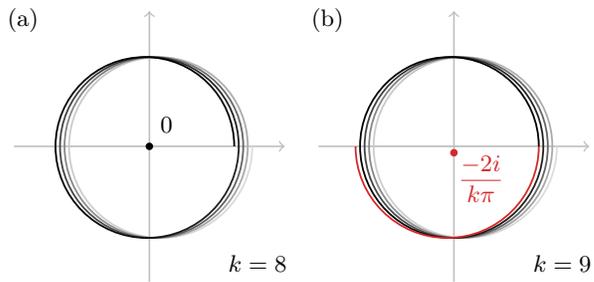
\begin{figure}[!t]
    \centering
    
    \begin{tikzpicture}[scale=1.2]
        \draw[gray!50,->,semithick] (-1.5,0) -- (1.5,0);
        \draw[gray!50,->,semithick] (0,-1.5) -- (0,1.5);
        \myshadetop{mygray2}{mygray3}{-0.90}{1}
        \myshadetop{mygray4}{mygray5}{-0.95}{1}
        \myshadetop{mygray6}{mygray7}{-1}{1}
        \myshadetop{mygray8}{mygray8}{-1.05}{1}
        \myshadebottom{mygray2}{mygray1}{-0.90}{1.025}
        \myshadebottom{mygray4}{mygray3}{-0.95}{1.025}
        \myshadebottom{mygray6}{mygray5}{-1}{1.025}
        \myshadebottom{mygray8}{mygray7}{-1.05}{1.025}
        \node at (0,0)[circle,fill,inner sep=1pt]{};
        \node at (0.19,0.24) {$0$};
        \node at (1.2,-1.3) {$k=8$};
        \node at (-1.4, 1.4) {(a)};
    \end{tikzpicture}
    \vspace{0.2cm}
    \begin{tikzpicture}[scale=1.2]
        \draw[gray!50,->,semithick] (-1.5,0) -- (1.5,0);
        \draw[gray!50,->,semithick] (0,-1.5) -- (0,1.5);
        \myshadetop{mygray2}{mygray3}{-0.90}{1}
        \myshadetop{mygray4}{mygray5}{-0.95}{1}
        \myshadetop{mygray6}{mygray7}{-1}{1}
        \myshadetop{mygray8}{mygray8}{-1.05}{1}
        \myshadebottom{mygray2}{mygray1}{-0.90}{1.025}
        \myshadebottom{mygray4}{mygray3}{-0.95}{1.025}
        \myshadebottom{mygray6}{mygray5}{-1}{1.025}
        \myshadebottom{mygray8}{mygray7}{-1.05}{1.025}
        \myshadebottom{myred}{myred}{-1.10}{1.025}
        \node at (0,-0.07073553)[circle,myred,fill,inner sep=1pt]{}; 
        \node[myred] at (0.30,-0.36) {$\dfrac{-2i}{k\pi}$};
        \node at (1.2,-1.3) {$k=9$};
        \node at (-1.4, 1.4) {(b)};
    \end{tikzpicture}
    
    \caption{\textbf{Integer moments} - (a) Even moments. The $k$th moment of $e^{-i \alpha_p/2}$ for even $k$ is zero since it corresponds to the barycenter of $k/2$ superposed unit circles. (b) Odd moments. The $k$th moment of $e^{-i \alpha_p/2}$ with $k$ odd is non-zero, as it corresponds to the barycenter of $(k-1)/2$ unit circles and one semi-circle (in red).}
    \label{fig:moment_computation}
\end{figure}

The quantity $\inlinebraketmatrix{x}{\overline{U_0^t}}{x}=\inlinebraketmatrix{x}{\overline{\left(U_x \, U_p\right)^t}}{x} $ can be expanded explicitly as a sum over paths by introducing a closure relation $\mathbb{1} = \sum_p \ketbra{p}$ for each step $t' \in \{1, \dots, t \}$. Since only $U_p$ is random,
it can be expressed in terms of correlators of the evolution operator $U_p$ as
\begin{equation}
\inlinebraketmatrix{x}{\overline{U_0^t}}{x} =\!\!\!\! \sum_{p_0, \dots p_t} \!\!\!\braket{x}{p_t} \prod_{j=1}^t \braketmatrix{p_j}{U_x}{p_{j-1}} \; \overline{ \prod_{j=0}^{t-1} U_{p_j}} \braket{p_0}{x}\,.
\label{eq:ballistic_decay_U0}
\end{equation}
Since the phases $\alpha_p$ are uncorrelated and uniformly distributed in $\left[ 0, 2 \pi \right]$ the quantities $e^{-i \alpha_p}$ are uniformly distributed over the unit circle and all the moments of $U_p$ are zero (see Fig.~\ref{fig:moment_computation}a). The overlap \eqref{eq:ballistic_decay_U0} then reduces to $\inlinebraketmatrix{x}{\overline{U_0^t}}{x}= \delta_{t, 0}$. The spectral function \eqref{eq:qkr_spectral_function_from_ballistic} is therefore constant with respect to the position, and so is the background \eqref{eq:qkr_background}. The same goes for the half-kick model with evolution operator $V_{1/2}$ because of Eq.~\eqref{eq:equiv_0_pos_nondiag}. The oscillations seen in Fig.~\ref{fig:qkr_evolution}a therefore correspond to interference effects. By contrast, for the $U_{1/2}$ model illustrated in Fig.~\ref{fig:qkr_evolution}b, the background itself is oscillatory, as we shall now see.

\begin{figure}[!t]
    \centering
    \includegraphics[width=\linewidth]{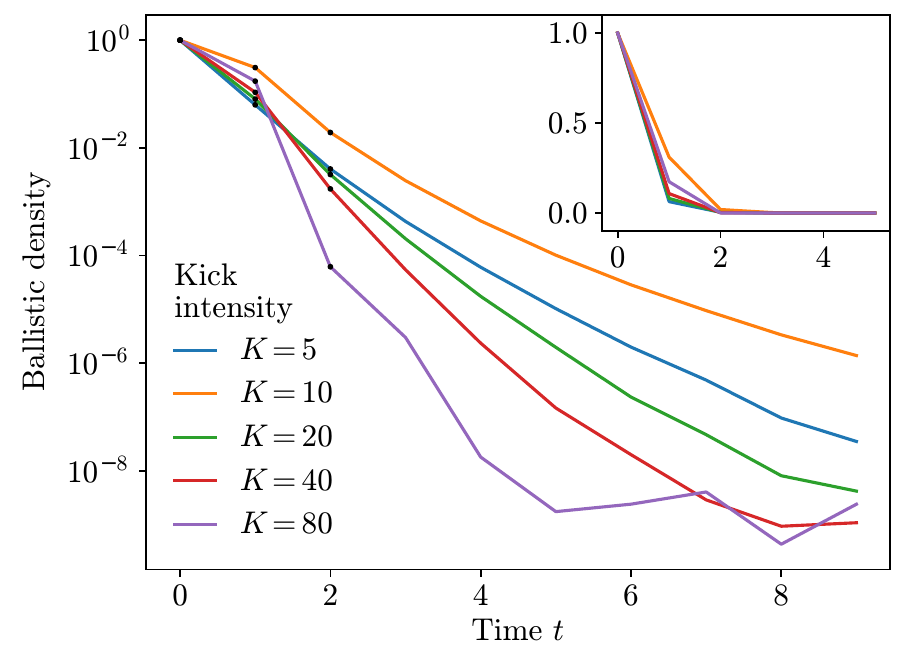}
    \caption{\textbf{Ballistic decay time} - Numerically computed ballistic density $n_b(x_0,t) = |\frac{1}{N}\bra{x_0} \overline{U^t_{1/2}} \ket{x_0}|^2$ for the $U_{1/2}$ model (full lines) in logarithmic scale. The parameters are $N = 800$, $\hbar_e = 2.89$, $x_0 = -\frac{\pi}{2}$ and $K \in \{5, 10, 20, 40, 80\}$. The disorder average is done over $10^6$ realizations. The black dots are obtained from the theoretical predictions at time $t=1$, Eq.~\eqref{eq:ballistic_decay_t1}, and at time $t=2$, Eq.~\eqref{eq:ballistic_decay_t2}. Inset: same as the main panel but in linear scale.}
    \label{fig:ballistic_decay_K}
\end{figure}

\subsection{Background for $U_{1/2}$}
\label{subsec:bgnd_half_period}

In this case, the computation is a bit more tricky since $U_{1/2}$ contains terms $e^{-i \alpha(\hat{p})/2}$ contributing diagonal random phases $e^{-ik\alpha_p/2}$ in momentum space with $k$ odd, which do not average to zero but to $-2i/(k\pi)$ (see Fig.~\ref{fig:moment_computation}b). The calculation steps are detailed in the Appendix \ref{app:background_computation}. Writing $\inlinebraketmatrix{x}{\overline{U_{1/2}^t}}{x}=\sum_{q\in\mathbb{Z}} e^{iqx} \, a_q(t)$, the Fourier series components are given by $a_0(t)=0$ and 
\begin{equation}
\label{eq:aqt}
a_q(t)= - \dfrac{4N i^q}{\pi^2} \sum_{m=0}^{\lfloor \frac{t-1}{2} \rfloor} g(m,t) J_0^{t-2m-1}   (-1)^{q m}J_{q}^{2m+1} 
\end{equation}
for $q\neq 0$, with 
\begin{equation}
 g(m,t) = \sum_{k=m}^{t-1-m} \, \frac{\binom{k}{m} \, \binom{t-k-1}{m}}{(2k+1)(2t-2k-1)}\,
   \label{eq:binom} 
\end{equation}
and Bessel functions $J_0$ and $J_q$ evaluated at $K/\hbar_e$.

For $t=1$ and $t=2$, this gives
\begin{align}
   \label{eq:ballistic_decay_t1}
   &\frac{1}{N}\inlinebraketmatrix{x}{\overline{U_{1/2}}}{x} = \dfrac{4}{\pi^2} \left[ J_0\left( K/\hbar_e \right) - e^{i (K/\hbar_e) \cos x} \right] \\
   &\frac{1}{N}\inlinebraketmatrix{x}{\overline{U^2_{1/2}}}{x} = \frac{2}{3} \, J_0(K/\hbar_e) \, \inlinebraketmatrix{x}{\overline{U_{1/2}}}{x}.
   \label{eq:ballistic_decay_t2}
   \end{align}
Both expressions are in very good agreement with numerical simulations, as illustrated by the black dots in Fig.~\ref{fig:ballistic_decay_K}.
\begin{figure}
    \centering
    \includegraphics[width=\linewidth]{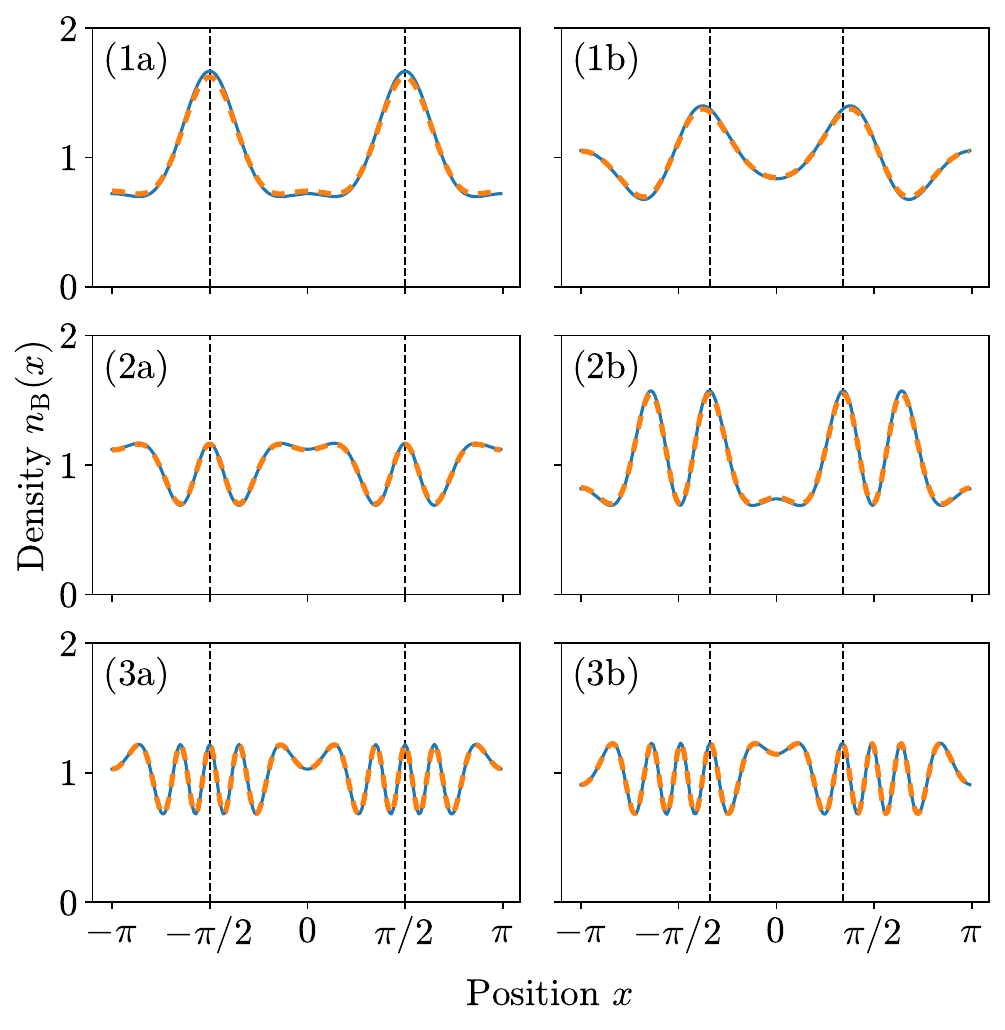}
       \caption{\textbf{Spatial oscillations of the background} - Numerical plot of the background (solid blue line) for the $U_{1/2}$ model with parameters $N = 200$, $\hbar_e = 2.89$, $K$ taking the values  $10$ for panels (1a,~1b), $20$ for panels (2a,~2b), and $40$ for panels (3a,~3b). The initial position is $x_0 = -\frac{\pi}{2}$ for panels (a) and $x_0 = -\frac{\pi}{3}$ for panels (b). The vertical dashed line mark the positions $\pm x_0$. The disorder average is done over 10~000 realizations. The approximate theoretical background given by Eq.~\eqref{eq:qkr_apprx_analytical_background} is displayed as an 
    orange dashed line and is barely distinguishable from the numerical calculations.}
    \label{fig:background_examples}
\end{figure}

Truncating the sum in Eq.~\eqref{eq:qkr_spectral_function_from_ballistic} to the first values of $t$ provides a fairly good approximation of the average spectral function $\bar{A}(x, \omega)$. A simple approximate expression for the spectral function can be obtained if we keep only the terms $|t|\leqslant 1$ in \eqref{eq:qkr_spectral_function_from_ballistic}. Using the explicit expression \eqref{eq:ballistic_decay_t1}, it reads 
\begin{equation}
    \bar{A}(x, \omega) \approx 1 + \frac{8}{\pi^2} \left[ \cos(\omega) J_0\left( \frac{K}{\hbar_e} \right) - \cos( \omega - \frac{K}{\hbar_e} \cos x ) \right].
\label{eq:qkr_apprx_analytical_spectral_function}
\end{equation}
Using this expression for the spectral function, we can derive approximate analytical formulas
for the density of states $\nu(\omega)$ using Eq.~\eqref{eq:qkr_normalization_A_nu_x} and for the background using Eq.~\eqref{eq:qkr_background}. We obtain a flat density of states, and a background given by
\begin{align}
    n_{\text{B}}(x, x_0) &\approx 1 + \frac{32}{\pi^4} \left\{ J_0^2\left( \frac{K}{\hbar_e} \right) \right.\nonumber\\
    &- J_0\left( \frac{K}{\hbar_e} \right) \left[ \cos\left( \frac{K}{\hbar_e} \cos x_0 \right) +  \cos\left( \frac{K}{\hbar_e} \cos x \right) \right] \nonumber\\
    &\left.+ \cos\left( \frac{K}{\hbar_e} \left( \cos x_0 - \cos x \right) \right) \right\}.
\label{eq:qkr_apprx_analytical_background}
\end{align}
This formula shows that we expect the amplitudes of spatial oscillations to be governed by $J_0(K/\hbar_e)$ and the typical scale of oscillations to be controlled by $\pi/\ell_s \sim \pi\hbar_e/K$.
The numerical results displayed in Fig.~\ref{fig:background_examples} show that this approximate formula is very accurate despite keeping only terms $|t|\leq 1$ in Eq.~\eqref{eq:qkr_spectral_function_from_ballistic}. Adding more terms slightly improves the approximation but, because of the exponential decay (see Fig.~\ref{fig:ballistic_decay_K}) of their amplitude with a characteristic time $\tau_C \sim 1$ for our range of parameters $K \in [5, 200]$ and $\hbar_e = 2.89$, there is no need to go beyond the first term here.
However, for small values of $K$, the characteristic time $\tau_C$ increases, and deviations are visible (see e.g.~Fig.~\ref{fig:background_examples}, panel 1a). When considering the terms associated with next values of $t$ we recover the agreement with the numerical simulations.

Note that the expressions we derived are obtained in the large-$N$ limit. As the plots in Fig.~\ref{fig:background_examples} show, already at $N=200$ the agreement of Eq.~\eqref{eq:qkr_apprx_analytical_background} with numerics is very good.

\subsection{CBS and CFS peaks}
\label{subsec:cs_cbs_cfs}

\begin{figure}[!t]
    \centering
    \includegraphics[width=\linewidth]{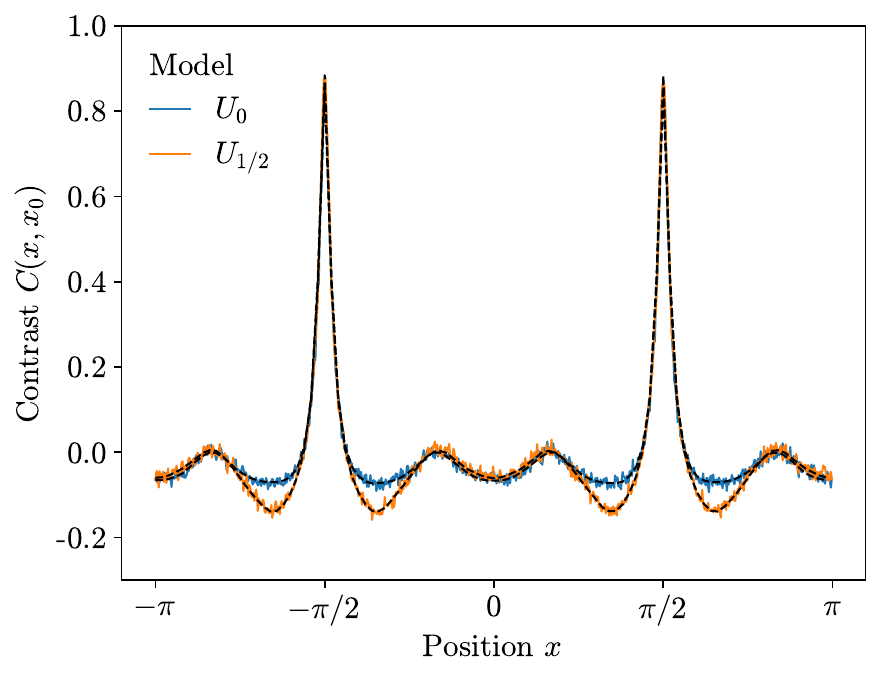}
    \caption{\textbf{Contrast profile at large time}. The contrast profile at large time $C(x,x_0)$ is given for the $U_0$ model (solid blue line) and for the $U_{1/2}$ model (solid orange line). Both models being time-reversal invariant, their profiles are even in $x$. The curves are obtained by looking at the density profile at $t=1000$ and then using Eq.~\ref{eq:qkr_contrast} with $n_B$ the theoretical background. The parameters used are $N=800$, $K=20$, $\hbar_e=2.89$ and the disorder average is taken over 10~000 realizations. As a reference, the contrast obtained from exact diagonalization of 5000 matrices of size 100 and using Eq.~\ref{eq:qkr_steady_state1} is also shown (black dashed lines).}
    \label{fig:qkr_cbs_cfs_interference}
\end{figure}

As defined in Eq.~\eqref{decompn}, the probability density $n(x, x_0, t)$ is a sum of a time-independent interference-free classical background component $n_B(x,x_0)$ and a purely time-dependent interferential component $n_I(x, x_0, t)$. The interference part of the contrast for the spatial profile is defined as
\begin{equation}
    C(x, x_0, t) = \dfrac{n_I(x, x_0, t)}{n_B(x, x_0)} = \dfrac{n(x, x_0, t)}{n_B(x, x_0)} - 1. \\
\label{eq:qkr_contrast}
\end{equation}
The CFS and CBS peak contrasts at time $t$ are defined as $C_{\text{CFS}}(x_0,t) = C(x_0,x_0,t)$ and $C_{\text{CBS}}(-x_0,t) = C(-x_0,x_0,t)$, respectively, and were investigated in \cite{lemarie_coherent_2017} for the RQKR model $U_0$. Whereas the CBS peak develops fully over a time scale of the order of a few scattering mean free times, the CFS peak develops over a longer time scale related to the localization time. Its infinite-time limit, $C_{\text{CFS}}(x_0) = C_{\text{CFS}}(x_0,t\to\infty)$, can be expressed in terms of the Floquet eigenvectors as
\begin{equation}
    C_{\text{CFS}}(x_0) =  \dfrac{\sum_n \overline{|\varphi_n(x_0)|^4}}{\sum_n \overline{|\varphi_n(x_0)|^2}^2} - 1.
\label{eq:qkr_cfs_contrat}
\end{equation}
In Fig.~\ref{fig:qkr_cbs_cfs_interference}, we plot the contrast at infinite time $C(x, x_0)$ for the $U_0$ and $U_{1/2}$ models. Despite the large differences in the background observed in Fig.~\ref{fig:qkr_evolution}, the interference patterns displayed in Fig.~\ref{fig:qkr_cbs_cfs_interference} are very similar. In particular, both the CFS and CBS peaks coincide in width and height. Remaining discrepancies can be observed outside the peak regions. 
This is because the phase terms $e^{i \alpha_p /2}$ with a non-zero average not only contribute to the background, they also slightly modify the interferential part of the density.

Equation~\eqref{eq:qkr_cfs_contrat} can be rewritten as $C_{\text{CFS}}(x_0) = \sum_n \gamma_n(x_0) \, p_n(x_0) -1$ where we have introduced the quantities
\begin{equation}
    \gamma_n(x_0)=\dfrac{\overline{|\varphi_n(x_0)|^4}}{\overline{|\varphi_n(x_0)|^2}^2}\, \qquad p_n(x_0) = \dfrac{\overline{|\varphi_n(x_0)|^2}^2}{\sum_n \overline{|\varphi_n(x_0)|^2}^2}\,.
    \label{eq:kurto_proba}
\end{equation} 
Writing the complex eigenfunction as $\varphi_n(x) = R_n(x) + i I_n(x)$ and assuming identical statistical properties and zero mean for the independent random variables $R_n$ and $I_n$, we arrive at $\gamma_n = (1+\Gamma_n)/2$ where $\Gamma_n = \overline{R^4_n}/\overline{R^2_n}^2$. Further assuming that the statistical properties of the eigenfunctions do not depend on $n$, we find:
\begin{equation}
    C_{\text{CFS}}(x_0) = \dfrac{\Gamma (x_0) -1}{2}\,.
    \label{eq:kurto_contrast}
\end{equation}
For eigenfunctions obeying random matrix statistics~\cite{kravtsov_random_2012}, the random variable $R_n$ obeys the Porter-Thomas distribution~\cite{falcao_2022} and we have, for $N \to \infty$, $\Gamma (x_0) = 3$. This leads to the prediction $C_{\text{CFS}}(x_0) =1$, in good agreement with Fig.~\ref{fig:qkr_cbs_cfs_interference}. The slight deviation observed could be explained by the strong localization of eigenfunctions~\cite{lee_dynamics_2014}. Indeed, with the parameters used, we have $\xi/\ell_s = K/(4\hbar_e) \sim 2$, and thus the localization length $\xi$ is of the order of a few scattering  mean free paths only, so that microscopic details become relevant.

\section{Discussion}
\label{sec:conclusion}

First implemented with a dilute sample of ultracold sodium atoms in a periodic standing wave~\cite{moore_1994, raizen_experimental_1995, moore1995atom}, the QKR model has become an effective experimental and theoretical tool to explore quantum transport phenomena in disordered media~\cite{garreau2017quantum, lemarie_2009, Hainaut_2019}. 
Among such effects stand the CBS and CFS peaks which are subtle interference effects resisting disorder average. Both peaks can provide experimental signatures for Anderson transitions or criticality~\cite{ghosh_cbs_2015, ghosh_cfs_2017, martinez_coherent_2021, martinez_coherent_2023}. If the CBS effect has been recently observed with cold atoms~\cite{labeyrie_cbs_2012, jendrzejewski_cbs_2012}, a direct observation of the CFS peak is still unpublished. However, as the present work illustrates, accessing the CBS or CFS contrast requires a careful handling of the background on top of which these interference patterns emerge. Indeed, we have shown that simple modifications of the paradigmatic quantum kicked rotor protocol can lead to highly nontrivial and strong spatial oscillations in the spatial density. We have further clarified that this oscillatory features mainly come from dramatic changes of the disorder-averaged spectral functions, thereby profoundly affecting the classical background itself at a spatial scale controlled by the inverse mean free path $\ell_s^{-1} \sim \hbar_e/K$. We have obtained an analytical description for this background that perfectly matches numerical calculations. For all models considered, we found that the interference component of the spatial density also exhibits some residual oscillations, albeit to a much lesser extent than the background itself.

So far we were not able to fully understand the remaining discrepancy between the interference part of the two models. since the eigenstates of $U_0$ and $U_{1/2}$ are related by the disorder-dependent unitary $\sqrt{U_p}$, we suspect that they relate to subtle and nontrivial correlations between these eigenstates, yet to be uncovered.

The value of the CFS contrast is directly affected by symmetries through eigenvector statistics. For the models considered here, the only relevant symmetry is time reversal, implying a theoretical CFS contrast equal to unity. However, if one considers symmetric phase disorder $\alpha_{-p}=\alpha_p$, the models becomes $\mathsf{P}$, $\mathsf{T}$ and $\mathsf{PT}$ symmetric. In this case, the CFS contrast reaches first a maximal value 3 over the time scale of the localization while the CBS contrast remains at unity. At very large times compared to the localization time, the CFS and CBS contrast finally equalize and reach the stationary value 2. We will detail these results in a forthcoming publication.
It is also possible to break the time-reversal invariance by changing the potential~\cite{lemarie_coherent_2017, thaha_symmetry_1993}. In this case, the CBS peak disappears but not the CFS peak. This was also observed for the Ruijsenaars-Schneider model~\cite{martinez_coherent_2023}. 

Our results easily generalize to arbitrary initial times $t_0$ of the kicking sequence as well as other kick strength ratios $r$. Because of Eq.~\eqref{eq:qkr_half_kick_evolution_operator}, the spatial density for $V_r$ and for $U_0$ remain identical and nothing special happens. For $t_0 \neq 1/2$, analytical calculations can be performed along the same lines as for $t_0=1/2$. The spectral functions and backgrounds remain nontrivial for reasons similar to the case $t_0=1/2$ that are finally essentially encompassed by Fig.~\ref{fig:moment_computation}.

\acknowledgments
This study was supported by research funding Grants No.~ANR-18-CE30-0017 and ANR-19-CE30-0013, by the ANR research grant ManyBodyNet No.~ANR-24-CE30-5851, the EUR grant NanoX No.~ANR-17-EURE-0009 in the framework of the “Programme des Investissements d’Avenir”, and by the Singapore Ministry of Education Academic Research Fund Tier I (WBS No.~R-144-000-437-114). We thank Calcul en Midi-Pyrénées (CALMIP) and the National Supercomputing Centre (NSCC) of Singapore for computational resources and assistance. H.~Thomas thanks J.~Gong for financial support.

\appendix

\section{Parity and time reversal symmetries}
\label{appconsequences}

We introduce the parity $\mathsf{P}$, time reversal $\mathsf{T}$ and $\mathsf{PT}$ symmetries for spinless systems as
\begin{equation}
\mathsf{P} \left\{
    \begin{aligned}
        x & \rightarrow -x \\
        p & \rightarrow -p \\
        t & \rightarrow t
    \end{aligned}
    \right.,\quad
    \mathsf{T} \left\{
    \begin{aligned}
        x & \rightarrow x \\
        p & \rightarrow -p \\
        t & \rightarrow -t
    \end{aligned}
    \right.,\quad
    \mathsf{PT}
    \left\{
    \begin{aligned}
        x & \rightarrow -x \\
        p & \rightarrow p \\
        t & \rightarrow -t
    \end{aligned}
    \right.
\label{eq:time_reversal_operator}
\end{equation}
$\mathsf{P}$ is linear while $\mathsf{T}$ and $\mathsf{PT}$ are anti-linear. It is interesting to note that $\mathsf{PT}$ is a time reversal operator where the role of position and momentum are exchanged.

Let us now consider a Hilbert space $\mathcal{H}$ and a basis $\mathcal{B}$ spanning $\mathcal{H}$. The complex conjugation operator $K_\mathcal{B}$ is defined by $K_\mathcal{B}\ket{a} = \ket{a}$, $\forall a\in\mathcal{B}$, and its anti-linear action on kets expanded on $\mathcal{B}$ is
\begin{equation}
    \ket{\varphi} = \sum_{a\in\mathcal{B}} \, c_a \, \ket{a}\,, \qquad K_\mathcal{B}\ket{\varphi} = \sum_{a\in\mathcal{B}} \, c^*_a \, \ket{a}.
\end{equation}
Using the position basis $\ket{x}$ and momentum basis $\ket{p}$, it is then easy to show that $\hat{K}_x$ realizes $\mathsf{T}$ while $\hat{K}_p$ realizes $\mathsf{PT}$ since 
\begin{align}
\hat{K}_x\ket{x}&=\ket{x}\,,\qquad \hat{K}_x\ket{p}=\ket{-p}\\
\hat{K}_p\ket{p}&=\ket{p}\,,\qquad \hat{K}_p\ket{x}=\ket{-x}
\label{opJ}
\end{align}
An operator $\hat{H}$ acting on $\mathcal{H}$ can be expressed in the $\ket{p}$ or $\ket{x}$ bases as
\begin{equation}
\label{hata}
    \hat{H}=\sum_{p,p'}H^P_{pp'}\ket{p}\bra{p'}=\sum_{x,x'}H^X_{xx'}\ket{x}\bra{x'}\,,
\end{equation}
where $H^P$ and $H^X$ are the Hermitian matrices representing $\hat{H}$ in the momentum and position basis, respectively. 
The system described by $\hat{H}$ is time-reversal invariant if there exists a time-reversal operator commuting with $\hat{H}$. Using  \eqref{hata}, we readily have
\begin{align}   
\label{xHx}
\hat{K}_x\hat{H}\hat{K}_x=\hat{H}
&\ \Leftrightarrow \  H^X \textrm{ real and } JH^PJ=(H^{P})^*\,, \\
\hat{K}_p\hat{H}\hat{K}_p=\hat{H}&\ \Leftrightarrow \  H^P \textrm{ real and } JH^XJ=(H^{X})^*\,,
\label{pHp}
\end{align}
where $J$ is the matrix with unit entries on the anti-diagonal and zeroes everywhere else (it represents the action of the parity operator $\mathsf{P}$).

In the present paper, as shown by Eq.~\eqref{eq:qkr_time_reversal_Utau}, the system under consideration is $\mathsf{PT}$-symmetric.
It implies that, if $\ket{\varphi_a}$ is an eigenstate of $\hat{H}$ with real eigenvalue $\lambda_a$ ($\hat{H}$ is Hermitian), then $\hat{K}_p\ket{\varphi_a}$ is also an eigenstate of $\hat{H}$ with same eigenvalue $\lambda_a$. If $\lambda_a$ is non-degenerate and eigenstates are normalized then $\hat{K}_p\ket{\varphi_a}=e^{i\alpha_a}\ket{\varphi_a}$ where $\alpha_a$ is some eigenstate-dependent phase. 
This phase can always be swept out by introducing the eigenstate $\ket{\phi_a} = e^{i\alpha_a/2} \ket{\varphi_a}$. With this eigenstate redefinition, one has $\hat{K}_p\ket{\phi_a}=\ket{\phi_a}$ and thus $\phi_a(p) = \langle p\ket{\phi_a}$ real. By Fourier transform this also gives $\phi_a^*(-x)=\phi_a(x)$. Throughout this paper, we assume that this eigenket gauge fixing procedure has been implemented when discussing time-reversal symmetric Hamiltonians.

\section{Normalization conventions}

\label{sec:appendix:normalization}
We took the convention that $p$ is integer and $x\in [-\pi,\pi[$ discretized by steps of size $2\pi/N$. We choose normalization and Fourier transform conventions in such a way that in the $N\to\infty$ limit all sums over $x$ go to well-defined integrals. Orthonormalization relations read
\begin{equation}
\braket{p}{p'}=\delta_{p,p'}\,, \quad
    \braket{x}{x'}=N \delta_{x,x'} \underset{N\rightarrow \infty}{\longrightarrow} 2\pi\delta(x-x').
\end{equation}
Closure relations are given by
\begin{equation}  
\label{clos}
\mathbb{1} =  \sum_{p} \ketbra{p}\,,\quad
       \mathbb{1} = \frac{1}{N} \sum_{x} \ketbra{x}  \underset{N\rightarrow \infty}{\longrightarrow} \int  \frac{\dd{x}}{2\pi} \ketbra{x}\,.
\end{equation}
Fourier transform between the two bases is defined as
\begin{align}
    \ket{x} &= \sum_{p} \e{-i x p} \ket{p} \\
    \ket{p} &= \frac{1}{N} \sum_{x} \e{i x p} \ket{x} \underset{N\rightarrow \infty}{\longrightarrow}  \int \frac{\dd{x}}{2\pi} \e{i x p} \ket{x} ,
    \label{ndotk}
\end{align}
which corresponds to the overlap
\begin{equation}
        \braket{p}{x}=\e{-i x p} .
\end{equation}
We choose eigenfunction normalization as
\begin{equation}
        \sum_{p} \vqty{\phi_n(p)}^2=1
\end{equation}
and (note the prefactor $1/N$)
\begin{equation}
     1=\frac{1}{N} \sum_x \vqty{\phi_n(x)}^2
     \underset{N\rightarrow \infty}{\longrightarrow}  \int \frac{\dd{x}}{2\pi} \vqty{\phi_n(x)}^2.
    \label{norm_ev_alpha}
\end{equation}
Eigenkets are normalized by $\braket{\phi_n}{\phi_{n'}}=\delta_{n,n'}$, and thus
\begin{equation}
\sum_n \vqty{\phi_n(x)}^2=1  \qquad  \forall x\,.  
\end{equation}
The density and spectral function defined in Eqs.~\eqref{eq:defnu} and \eqref{eq:qkr_spectral_function} are normalized by
\begin{align}
     &\int_{-\pi}^\pi \text{d} \omega \, \nu (\omega) = 1, \label{eq:qkr_normalization_nu_omega} \\
   &\int_{-\pi}^\pi \text{d} \omega \, \bar{A}(x, \omega) = 1. \label{eq:qkr_normalization_A_omega}
\end{align}
Normalization in $x$ gives 
\begin{align}    
 \nu(\omega) =\frac{1}{N} \sum_{x} \bar{A}(x,\omega) \underset{N\rightarrow \infty}{\longrightarrow}  \int \frac{\dd{x}}{2\pi}\bar{A}(x,\omega).
    \label{eq:qkr_normalization_A_nu_x}
\end{align}
Finally, in going from discrete summations over eigenstates to integrals over quasi-energies, we have used the prescription
\begin{equation}
    \delta_{n,m} \to \dfrac{\delta (\omega_n-\omega_m)}{N \nu (\omega_n)}
\end{equation}
to establish Eq.~\eqref{eq:qkr_steady_state1}. It is valid in the bulk limit $N\to\infty$ and after disorder average~\cite{lee_dynamics_2014}.

\section{Analytic computation of the background}
\label{app:background_computation}
We consider the disorder-averaged propagation amplitude from initial position state $\ket{x_0}$ to position state $\ket{x}$:
\begin{equation}
    \inlinebraketmatrix{x}{\overline{U_{1/2}^t}}{x_0} = \braketmatrix{x}{\overline{\left(e^{-\frac{i \alpha(\hat{p})}{2}} \, e^{i \frac{K}{\hbar_e} \cos \hat{x}} \, e^{-\frac{i \alpha(\hat{p})}{2}}\right)^t}}{x_0}
    \label{eq:propa_amp} 
    \end{equation}
    which, inserting closure relations \eqref{clos} at each time step, becomes
 \begin{align}
&\inlinebraketmatrix{x}{\overline{U_{1/2}^t}}{x_0} = \sum_{p_0, \dots p_t} \braket{x}{p_t} \prod_{j=1}^t \braketmatrix{p_j}{e^{i \frac{K}{\hbar_e} \cos \hat{x}}}{p_{j-1}} \; \nonumber\\
&\hspace{1cm} \times\overline{e^{-\frac{i \alpha_{p_t}}{2}} ( \prod_{j=1}^{t-1} e^{-i \alpha_{p_j}} ) e^{-\frac{i \alpha_{p_0}}{2}}} \braket{p_0}{x_0}\,.
    \label{eq:ballistic_decay_calculation_2}
\end{align}
In \eqref{eq:ballistic_decay_calculation_2}, the only paths $(p_0, \dots, p_t)$ which do not average to zero are the ones for which the set $\{p_1,...,p_{t-1}\}$ contains $k$ momenta equal to $p_t$ and  $(t-1-k)$ equal to $p_0$ (see Fig.~\ref{fig:moment_computation}), with $p_0\neq p_t$. Integer moments of order $k$ of  $e^{-i \alpha_p/2}$ for $k$ odd are given by $\overline{e^{- i k \alpha_p/2}} = -2 i/(k\pi)$, and thus Eq.~\eqref{eq:ballistic_decay_calculation_2} becomes
\begin{align}
\label{eq:ballistic_decay_calculation_4}
   \inlinebraketmatrix{x}{\overline{U_{1/2}^t}}{x_0}   &= - \dfrac{4}{\pi^2}\sum_{p_t \neq p_0} \sum_{k=0}^{t-1} f(k,t) \\
   &\times\!\!\!\!\!\sum_{(p_1, \dots, p_{t-1})}\!\!\!\!\!\! \braket{x}{p_t}
\prod_{j=1}^t \braketmatrix{p_j}{e^{i \frac{K}{\hbar_e} \cos \hat{x}}}{p_{j-1}} \braket{p_0}{x_0}\,, \nonumber\\
f(k,t)&=\frac{1}{(2k+1)(2t-2k-1)}\,,
\end{align}
where the sum $\sum_{(p_1, \dots, p_{t-1})}$ is restricted to momenta $\left(p_i\right)_{1\leqslant i \leqslant t-1}$ such that $k$ of them are equal to $p_t$ and the remaining $(t-1-k)$ ones to $p_0$.

Setting $q_j=p_j-p_{j-1}$ the momentum transfer at each step, we have 
$\inlinebraketmatrix{p_j}{e^{i \frac{K}{\hbar_e} \cos \hat{x}}}{p_{j-1}}=i^{q_j}J_{q_j}(K/\hbar_e)$.
Since momenta $\left(p_i\right)_{1\leqslant i \leqslant t-1}$ only take values $p_t$ or $p_0$, the $q_j$ can only take the values $q_j=0,\pm q$, where $q=p_t-p_0$. As a consequence, Eq.~\eqref{eq:ballistic_decay_calculation_4} reduces to
\begin{align}
\label{eq:ballistic_decay_calculation_5}
  & \inlinebraketmatrix{x}{\overline{U_{1/2}^t}}{x_0}   = - \dfrac{4}{\pi^2}\sum_{p_t \neq p_0} \sum_{k=0}^{t-1} f(k,t) e^{i(p_t x-p_0 x_0)} \\
   &\times \!\!\!\!\!\!\sum_{N_0,N_1,N_2}  \!\!\sum_qi^q (-1)^{q N_2}J_0^{N_0}J_{q}^{N_1+N_2}\mathcal{N}_k(N_0,N_1,N_2)\delta_{q,p_t-p_0}
   \,\nonumber
\end{align}
(we used the fact that $J_{-q}=(-1)^q J_q$ for integer $q$).
Here $N_0$ (resp.~$N_1$ and $N_2$) is the number of occurrences of zero (resp.~$q$ and $-q$) momentum transfers, and $\mathcal{N}_k(N_0,N_1,N_2)$ is the number of configurations $(p_1,...,p_{t-1})$ of entries with $k$ times $p_0$ and $t-1-k$ times $p_t$ giving the momentum transfer occurences $N_0,N_1,N_2$. Bessel functions $J_0$ and $J_q$ are evaluated at $K/\hbar_e$. In \eqref{eq:ballistic_decay_calculation_5} one can perform the summation over $p_0$ and $p_t$; one gets
\begin{equation}
   \sum_{p_t \neq p_0} e^{i(p_t x-p_0 x_0)} \delta_{q,p_t-p_0}=N e^{i x q} \delta_{x,x_0} 
\end{equation}
for $q\neq 0$, and 0 if $q=0$, so that \eqref{eq:ballistic_decay_calculation_5} becomes
\begin{align}
\label{eq:ballistic_decay_calculation_6}
   \inlinebraketmatrix{x}{\overline{U_{1/2}^t}}{x_0}   &= - \dfrac{4N}{\pi^2} \sum_{k=0}^{t-1} f(k,t)\!\!\!\!\!\!\sum_{N_0,N_1,N_2}  \!\!  J_0^{N_0} \mathcal{N}_k(N_0,N_1,N_2) \nonumber \\
   &\times\sum_{q\neq 0} i^q (-1)^{q N_2}J_{q}^{N_1+N_2}e^{i x q} \delta_{x,x_0} 
   \,.
\end{align}
This shows in particular that $\inlinebraketmatrix{x}{\overline{U_{1/2}^t}}{x_0} = \inlinebraketmatrix{x}{\overline{U_{1/2}^t}}{x}\, \delta_{x,x_0}$: the ballistic contribution vanishes outside the forward peak.

In order to find the combinatorial coefficient $\mathcal{N}_k(N_0,N_1,N_2)$, we note the formal analogy of the problem with a one-dimensional $A-B$ binary alloy model. By associating symbol $A$ to momenta $p_t$ and symbol B to momenta $p_0$, the combinatorics amounts to determining the number of configurations $\mathcal{N}(N_1,N_2,t,k)$ of a finite chain $A-\mathcal{L}-B$ where the subchain $\mathcal{L}$ has length $(t-1)$ with $k$ symbols $A$, and contains $N_1$ bonds $A-B$ and $N_2$ bonds $B-A$.
Since the extremities of the chain are different, $M=N_1+N_2$ must be odd, with $1\leq M \leq t$; we set $M=2m+1$. Moreover, since the chain starts with $A$ and ends with $B$, one must have $N_1-1=N_2=m$, with $0\leq m \leq \lfloor \frac{t-1}{2} \rfloor$ and of course $N_0=t-2m-1$. The number of configurations reads
\begin{equation}
   \mathcal{N}_k(N_0,N_1,N_2) = \binom{k}{m} \binom{t-1-k}{m}\,
\end{equation}
and \eqref{eq:ballistic_decay_calculation_6} yields
\begin{align}
\label{eq:ballistic_decay_calculation_7}
  & \inlinebraketmatrix{x}{\overline{U_{1/2}^t}}{x_0}   = - \dfrac{4N}{\pi^2} \sum_{k=0}^{t-1} f(k,t) \sum_{m=0}^{\lfloor \frac{t-1}{2} \rfloor} J_0^{t-2m-1}   \\
   &\quad\times\binom{k}{m} \binom{t-1-k}{m}\sum_{q\neq 0} i^q (-1)^{q m}J_{q}^{2m+1}e^{i x q} \delta_{x,x_0} 
   \,.\nonumber
\end{align}
A straightforward calculation then leads to Eq.~\eqref{eq:aqt} supplemented by Eq.~\eqref{eq:binom}. For $t=1$, only the term $(m,k)=(0,0)$ contributes to the sums; using the identity
\begin{equation}
    \sum_q i^q J_q(z)e^{i q \theta}=e^{i z \theta},
\end{equation}
we get Eq.~\eqref{eq:ballistic_decay_t1}.
For $t=2$, the terms $(m,k)=(0,0)$ and $(0,1)$ contribute, leading to Eq.~\eqref{eq:ballistic_decay_t2}. Subsequent terms are easily computed too but we found that the background computed with Eq.~\eqref{eq:qkr_background} and Eq.~\eqref{eq:qkr_spectral_function_from_ballistic} restricted to $|t| \leq 1$ is already a good approximation.

\bibliography{biblio}

\end{document}